\documentclass[twocolumn,prb,showpacs,multicol,amsmath,amssymb]{revtex4}
\usepackage[dvips]{graphicx}
\usepackage{graphicx}
\usepackage{dcolumn}
\usepackage{bm}
\usepackage{graphics}
\usepackage{epsfig,color}

\newcommand{\be}{\begin{equation}}
\newcommand{\ee}{\end{equation}}
\newcommand{\bea}{\begin{eqnarray}}
\newcommand{\eea}{\end{eqnarray}}

\newcommand{\bS}{\bf S}

\begin{document}
\title{Magnetic Entanglement in Spin-1/2 XY Chains}

\author{Fatemeh Khastehdel Fumani$^{1}$, Somayyeh Nemati$^{2}$, Saeed Mahdavifar$^{3}$, Amirhosein Darooneh$^{2}$}
\affiliation{ $^{1}$Department of Basic Sciences, Langaroud branch,Islamic Azad University, Langaroud, Iran}
\affiliation{ $^{2}$Department of Physics, University of Zanjan, 45196-313, Zanjan, Iran}
\affiliation{ $^{3}$Department of Physics, University of Guilan, 41335-1914, Rasht, Iran}

\date{\today}

\begin{abstract}

  In the study of entanglement in a spin chain, people often consider the nearest-neighbor spins. The motivation is the prevailing role of the short range interactions in creating quantum correlation between the 1st neighbor (1N) spins. Here, we address the same question between farther neighbor spins.
  We consider the one-dimensional (1D) spin-1/2 XY model in a magnetic field. Using the fermionization approach, we diagonalize the Hamiltonian of the system. Then, we provide the analytical results for entanglement between the 2nd, 3rd and 4th neighbor (denoted as 2N, 3N, and 4N respectively) spins. We find a magnetic entanglement that starts from a critical entangled-field ($h_c^{E}$) at zero temperature. The critical entangled-field depends on the distance between the spins. In addition to the analytical results, the mentioned phenomenon is confirmed by the numerical Lanczos calculations. By adding the temperature to the model, the magnetic entanglement remains stable up to a critical temperature, $T_c$. Our results show that entanglement spreads step by step to farther neighbors in the spin chain by reducing temperature. At first, the 1N spins are entangled and then further neighbors will be entangled respectively. $T_c$ depends on the value of the magnetic field  and will be maximized at the quantum critical field.

 \end{abstract}
\maketitle


\section{Introduction}\label{sec1}

The quantification and control of quantum entanglement, for quantum information processing purposes through teleportation\cite{bennet93, bouwmeester97, gorbachev00} and superdense coding\cite{bennet92, hao01}, have derived extensive experimental\cite{bennet96, hill97, wotters98} and theoretical\cite{vedral97,horodecki98, rains99} researches in the recent decades. Physically interesting in this area, solid state systems also have been used as a topic of many studies for their potential applications on transmitting a quantum state by using entanglement\cite{bayat09}. Particularly one-dimensional quantum spin systems comprise many nonclassical properties that make them interesting for studying spin entanglement.
The magnetic behavior of these systems is explained through the isotropic Heisenberg model or its special cases such as the Ising, XY, and XXZ models. In addition to many compounds with common isotropic Heisenberg structure like $La_{2}CuO_{4}$\cite{kastner98}, $CuGeO_3$\cite{hase93}, and $LiCuVO_{4}$\cite{gibson04}, there are also some experimental examples such as $Fe(C_5H_5NO)_6(ClO_4)_2$\cite{algra78} and $Cs_2CoCl_4$\cite{blote75} which are well described by Ising and XY models respectively.
It has been found that the systems with XY interaction can be used as quantum dots\cite{kwek09, loss98, imamoglu99}. Therefore, many studies have been devoted to the exploration of important features of this model\cite{imamoglu99, loss98, raussendrof01,kwek09, wang01}. Also the fact that its eigenvalues can be exactly solved through the Jordan-Wigner (JW) transformation makes this model very practical\cite{jordan28}.

The zero-temperature quantum behavior of the spin-1/2 isotropic antiferromagnetic XY chain shows that it is in the Luttinger Liquid (LL) phase and undergoes a quantum phase transition by applying an external magnetic field\cite{takahashi99}. In the LL regime, the 1N spins are entangled\cite{Amico02}. Increasing the magnetic field reduces the entanglement measure until the spins become completely disentangled in the magnetic fields larger than the quantum critical field, $h_{c}$.

The thermal entanglement (TE) is of particular interest and demonstrates that the non-local correlations persist even in the thermodynamic limit\cite{Nielsen98, Arnesen01}. Studying the TE, zero-temperature entanglement, and the relation between the entanglement and the quantum phase transitions may lead to a relationship between the quantum information theory and the condensed matter physics.

Since the TE can be experimentally verified via measurable macroscopic parameters\cite{Ghosh03, Vedral03, Wiesniak05,fukuhara15}, many efforts have been dedicated to quantify it. Also, it has been recently demonstrated that an entangled pair of spins separated by several lattice spacing within an antiferromagnetic chain can be a suitable candidate for application in the quantum information processing\cite{sahling15}. This work also highlights the importance of studying the TE in different distances.

According to Ref.~[\onlinecite{shu09}], TE of the 1N spins decreases by increasing the temperature. There is a critical temperature $(T_{c})$ where TE will be zero\cite{shu09}. It was found that the critical temperature is independent of the magnetic field. This property is not exclusive to the XY model, but it has been observed in other 1D spin-1/2 systems as well\cite{ arnesen01, Mehran14}. In addition to the 1N spins, the zero-temperature entanglement between two spins at arbitrary distances was also studied using analytical and numerical calculations in the spin-1/2 XY chains\cite{nishimori05,osborne02}. It has been shown that at zero temperature, the entanglement between the non-nearest neighbor spins becomes finite around the quantum critical field.
The entanglement remains finite even if the distance between the spin pair reaches the system size\cite{nishimori05}. The concurrence and the quantum discord for the 2N spins were also studied in the Ising and the XXZ spin chains\cite{Dillenschneider08}. It was found that the quantum correlations increases in the region close to the critical points. The scaling behavior of the quantum correlations between a spin pair at an farther distances in the XY chain has been investigated recently\cite{Maziero10,huang14}. It is shown that the quantum discord between the spin pairs positioned at distances further than two lattice spacing reveals a quantum phase transition. Moreover, the quantum discord may increase both with the temperature and the magnetic field in certain regions of the parameter space.

In this paper, we study an infinite 1D spin-1/2 isotropic XY model in the presence of a magnetic field. Using the JW transformation, we find an analytical solution for the TE between the 2N, 3N, and 4N spin pairs in the thermodynamic limit. In absence of the magnetic field, the non-nearest neighbor pairs are not entangled. They remain unentangled until the magnetic field reaches the critical value $h_{c}^{E}$ which depends on the distance between the spins. We show that the $h_{c}^{E}$ is a function of $h_c$. By further increasing the magnetic field, the entanglement between the 2N, 3N, and 4N spin pairs increases to a maximum value before settling to zero at the quantum critical field, $h_c$. By taking the temperature into account, the concurrence within 2N, 3N, and 4N pairs decreases and reaches zero at a magnetic field dependent on critical value $T_c$. Note that this critical temperature was found to be field-independent for the 1N spins\cite{shu09, Mehran14}. $T_c$ is maximum at the quantum critical field value. For the magnetic fields larger than the quantum critical value, two critical temperatures can be found.

This paper is organized as follows. In the next section we introduce the model and the Hamiltonian which is diagonalized by using the JW transformation. In section \ref{sec3}, we discuss our results on the thermal behavior of the concurrence  between the 2N, 3N, and 4N spins. In section \ref{sec4} a conclusion and the summary of the results is presented.

\section{THE MODEL}\label{sec2}

The Hamiltonian of a 1D spin-1/2 isotropic XY model in a external magnetic field is written as
\begin{eqnarray}
{\cal H} &=& J \sum_{j=1}^{N} ({\bS}^{x}_{j}
{\bS}^{x}_{j+1}+{\bS}^{y}_{j}
{\bS}^{y}_{j+1}) - h  \sum_{j=1}^{N}
{\bS}^{z}_{j},   \label{Hamiltonian}
\end{eqnarray}
where $\bS_{j}$ is the spin-1/2 operator of the $j$-th site. $J>0$ denotes the antiferromagnetic exchange coupling and $h$ is the magnetic field.
We use the JW transformation to diagonalize the Hamiltonian. It converts the spin operators into spinless fermion operators as below:

\begin{eqnarray}
{\bS}^{+}_{j}&=& a_{j}^{\dag}(e^{i\pi \sum_{l<j} a_{l}^{\dag}a_{l}}),\\
{\bS}^{-}_{j}&=& (e^{-i\pi \sum_{l<j} a_{l}^{\dag}a_{l}})a_{j},\\
{\bS}^{z}_{j}&=& a_{j}^{\dag}a_{j}-\frac{1}{2}. \label{fermion operators}
\end{eqnarray}
\\By performing this transformation, the Hamiltonian is mapped onto the Hamiltonian of a 1D non-interacting fermionic system,
\begin{eqnarray}
{\cal H_{\textit{f}}} &=& \frac{Nh}{2}+ J \sum_{j} (a_{j}^{\dag}a_{j+1}+
 a_{j}a_{j+1}^{\dag}) - h  \sum_{j} a_{j}^{\dag}a_{j}.  \label{fermionic Hamiltonian}
\end{eqnarray}
Using the Fourier transformation $a_{j} = \frac{1}{\sqrt{N}}\sum_{k}e^{-ikj} a_{k}$, the momentum space Hamiltonian is diagonalized as
\begin{eqnarray}
{\cal H_{\textit{f}}} &=&  \sum_{k} \varepsilon(k) a_{k}^{\dag}a_{k}.  \label{diagonalized Hamiltonian}
\end{eqnarray}

$\varepsilon(k)$ is the dispersion relation:
\begin{eqnarray}
\varepsilon(k) &=&  J \cos(k)- h. \  \label{diagonalized Hamiltonian}
\end{eqnarray}

\section{Thermal entanglement}\label{sec3}

We focus on the entanglement between two sites, which is quantified by the concurrence. As defined below the concurrence measures the non-local quantumness of the correlations\cite{bennet96, hill97, wootters98}.

\begin{figure}
\centerline{\psfig{file=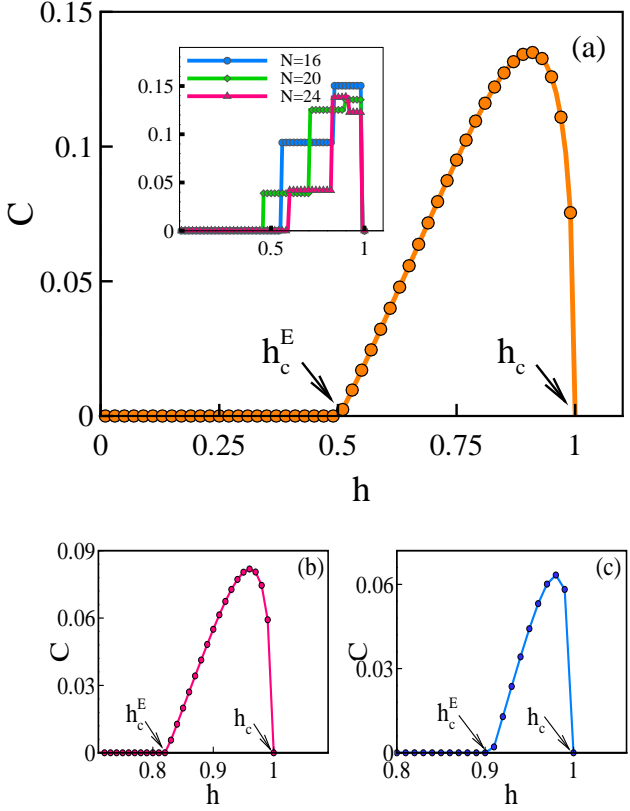,width=3.25in}}
\caption{(color online). The concurrence between (a) 2N, (b) 3N, and (c) 4N spins versus the applied magnetic field at zero temperature. The inset of (a) shows Lanczos results for the 2N spins obtained from the finite chains with lengths $N=16,20,24$ and exchange coupling $J=1$.}
\label{zero temperature}
\end{figure}

\begin{eqnarray}
C(\rho_{ij}) &=&  \max{\{0, \lambda_{1}-\lambda_{2}-\lambda_{3}-\lambda_{4}\}},  \label{C1}
\end{eqnarray}
 where $\lambda_{i}$s are square roots of the eigenvalues of the product matrix $R=\rho\widetilde{\rho}$ where $\widetilde{\rho}_{ij} =  (\sigma^{y}_{i} \otimes \sigma^{y}_{j}) \rho^{*}_{ij}
(\sigma^{y}_{i} \otimes \sigma^{y}_{j})$ and $\rho_{ij}$ is the reduced density matrix related to any (i,j) pair in the chain. $\lambda_i$s are real and non-negative even though $R$ is not necessarily Hermitian\cite{amico03}. The reduced density matrix contains all information about the involved spins. The density matrix in the standard basis, $\{ |\uparrow\uparrow\rangle, |\uparrow\downarrow\rangle, |\downarrow\uparrow\rangle, |\downarrow\downarrow\rangle \}$, can be expressed as:
{\small
\begin{eqnarray}
\rho_{ij}= \left(
             \begin{array}{cccc}
               <P_{i}^{\uparrow}P_{j}^{\uparrow}> & <P_{i}^{\uparrow}{\bS}_{j}^{-}> & <{\bS}_{i}^{-}P_{j}^{\uparrow}> & <{\bS}_{i}^{-}{\bS}_{j}^{-}> \\
               <P_{i}^{\uparrow}{\bS}_{j}^{+}> & <P_{i}^{\uparrow}P_{j}^{\downarrow}> & <{\bS}_{i}^{-}{\bS}_{j}^{+}> & <{\bS}_{i}^{-}P_{j}^{\downarrow}> \\
               <{\bS}_{i}^{+}P_{j}^{\uparrow}> & <{\bS}_{i}^{+}{\bS}_{j}^{-}> & <P_{i}^{\downarrow}P_{j}^{\uparrow}> & <P_{i}^{\downarrow}{\bS}_{j}^{-}> \\
               <{\bS}_{i}^{+}{\bS}_{j}^{+}> & <{\bS}_{i}^{+}P_{j}^{\downarrow}> & <P_{i}^{\downarrow}{\bS}_{j}^{+}> & <P_{i}^{\downarrow}P_{j}^{\downarrow}> \\
             \end{array}
           \right). \label{density matrix1}
\end{eqnarray}
}
The brackets symbolize the ground state and the thermodynamic average values at zero and a finite temperature, respectively. $P^{\uparrow}= \frac{1}{2}+{\bS}^{z}, P^{\downarrow}= \frac{1}{2}-{\bS}^{z}$, and ${\bS}^{\pm}= {\bS}^{x}\pm i{\bS}^{y}$. Following  the symmetry properties of the Hamiltonian, the density matrix must be real and symmetrical\cite{osterloh02}. It is found that only some elements of the density matrix are non-zero\cite{nag11, syljuasen03}:

\begin{eqnarray}
\rho_{ij}=\left(
            \begin{array}{cccc}
              X_{ij}^{+} & 0 & 0 &0 \\
              0 & Y_{ij}^{+} & Z_{ij}^{*} & 0 \\
              0 & Z_{ij} & Y_{ij}^{-} & 0 \\
              0 & 0 & 0 & X_{ij}^{-} \\
            \end{array}
          \right).
\label{density matrix2}
\end{eqnarray}
By considering the spin pair $(i,j)$ at the distance $m$ from each other, $j=i+m$, the density matrix elements can be obtained as:

\begin{eqnarray}
X_{i,i+m}^{+}&=& <n_{i}n_{i+m}>,\nonumber \\
Y_{i,i+m}^{+}&=& <n_{i}(1-n_{i+m})>,\nonumber \\
Y_{i,i+m}^{-}&=& <n_{i+m}(1-n_{i})>,\nonumber \\
Z_{i,i+m}&=& <a_{i}^{\dag}(1-2a_{i}^{\dag}a_{i})(1-2a_{i+1}^{\dag}a_{i+1})\nonumber \\
&\cdots&(1-2a_{i+m-1}^{\dag}a_{i+m-1})a_{i+m}>,\nonumber \\
X_{i,i+m}^{-}&=& <1-n_{i}-n_{i+m}+n_{i}n_{i+m}>,\  \label{elements}
\end{eqnarray}
where $n_{i}= a_{i}^{\dag}a_{i}$ is the occupation number operator. Therefore, the concurrence between this spin pair is:

\begin{eqnarray}
C(\rho) &=&  \max\{0,2(|Z_{i,i+m}|-\sqrt{X_{i,i+m}^{+}X_{i,i+m}^{-}})\}. \  \label{C2}
\end{eqnarray}


The $m=1$ case has been discussed in Ref.~[\onlinecite{shu09}]. Gong and Su have derived an equation for a unique $T_c$ so that the 1N spins are entangled at temperatures lower than $T_c$. Here we study the concurrence of the 2N, 3N, and 4N spins only. For these cases $Z_{i,i+m}$ and $X_{i,i+m}^{+}$ are obtained as follows:
\begin{eqnarray}
Z_{i,i+2}&=&f_2-2 f_0 f_2+2 f_1^{2}, \nonumber\\
X_{i,i+2}^{+}&=&f_{0}^{2} - f_{2}^{2},
\label{zx2}
\end{eqnarray}
\begin{eqnarray}
Z_{i,i+3}&=&4 (f_{1}^{3}-2 f_0 f_1 f_2+f_{2}^{2} f_1+f_{0}^{2} f_3 \nonumber\\
&-&f_{1}^{2} f_3+f_1 f_2-f_0 f_3)+f_3,\nonumber\\
X_{i,i+3}^{+}&=&f_{0}^{2} - f_{3}^{2},
\label{zx3}
\end{eqnarray}
\begin{eqnarray}
Z_{i,i+4}&=&8 (f_{1}^{4}-3 f_0 f_{1}^{2} f_2+2f_{1}^{2} f_{2}^{2}+2f_{0}^{2} f_1f_3 \nonumber\\
&+&f_{0}^{2} f_{2}^{2}-f_{2}^{4}-2f_0f_1f_2f_3+2f_1f_{2}^{2}f_3-2f_{1}^{3}f_3\nonumber\\
&+&f_{1}^{2}f_{3}^{2}-f_0f_2f_{3}^{2}-f_{0}^{3}f_4+2f_0f_{1}^{2}f_4-2f_{1}^{2}f_2f_4\nonumber\\
&+&f_0f_{2}^{2}f_4)+4(3f_{1}^{2}f_2-2f_0f_{2}^{2}-4f_0f_1f_3\nonumber\\
&+&2f_1f_2f_3+3f_{0}^{2}f_4-2f_{1}^{2}f_4+f_2f_{3}^{2}-f_{2}^{2}f_4)\nonumber\\
&+&2(2f_1f_3-3f_0f_4+f_{2}^{2})+f_4,\nonumber\\
\nonumber\\
X_{i,i+4}^{+}&=&f_{0}^{2} - f_{4}^{2},
\label{zx4}
\end{eqnarray}
where for the non-negative integer number $\textit{n}$


\begin{eqnarray}
f_{\textit{n}}=\frac{1}{2\pi}\int_{-\pi}^{\pi}e^{-ik\textit{n}} f(k)dk \label{Z1}.
\end{eqnarray}

\begin{figure}
\centerline{\psfig{file=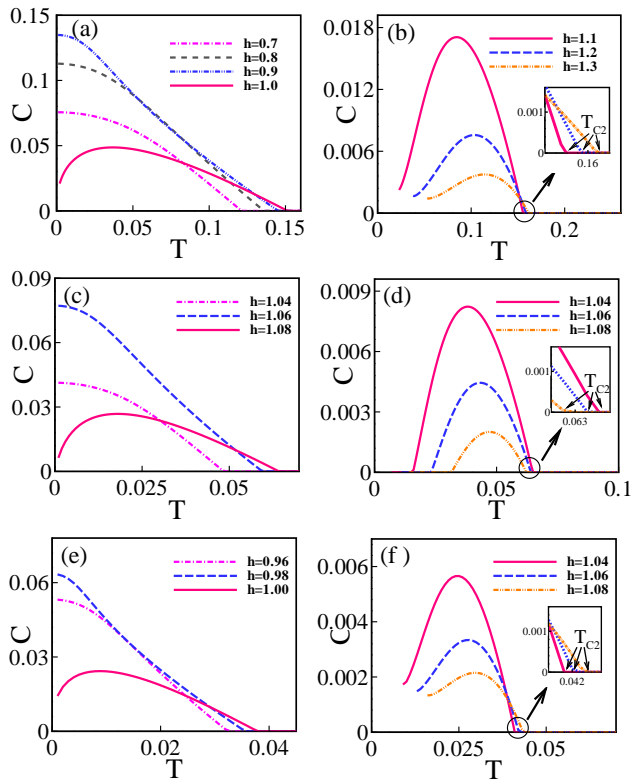,width=3.25in}}
\caption{(color online). The temperature dependence of the concurrence between (a) and (b) 2N, (c) and (d) 3N, and (e) and (f) 4N spins. The left-hand (right-hand) figures are for magnetic field values less (larger) than the quantum critical field.}
\label{hc}
\end{figure}

$f(k)=\frac{1}{1+e^{\beta \varepsilon(k)}}$ is the Fermi distribution function, where $\beta=\frac{1}{k_{B} T}$ and the Boltzmann constant $k_B=1$. Eq.~(\ref{zx2}), Eq.~(\ref{zx3}), and Eq.~(\ref{zx4}) are obtained using the Wick's theorem\cite{lieb61}.

In a system with a fixed exchange coupling, these integrals depend on two control parameters, i.e. temperature ($T$) and magnetic field ($h$). A particular interesting case is study of the quantum correlation between the 2N, 3N, and 4N spins at zero temperature. Fig.\ref{zero temperature} depicts the analytically calculated concurrence as a function of the magnetic field. Note that although the 2N, 3N, and 4N spin pairs follow the same trend in Figs.\ref{zero temperature}(a), (b), and (c) the concurrence measure decreases by increasing the distance. Also none of the pairs is entangled in the absence of the magnetic field. They remain unentangled up to a critical entangled-field, $h_c^{E}$. Fig.\ref{zero temperature} shows that $h_c^{E}$ for the 4N spins is closer to the quantum critical field value than the other two spin pairs. It is notable that there is a relationship between $h_c^{E}$ and $h_c$. Our results show that the critical entangled-field is $\frac{1}{2}h_c$ and approximately $\frac{4}{5}h_c$ and $\frac{9}{10}h_c$ for the 2N, 3N, and 4N pairs respectively. Based on these results, one may propose a general relation for $h_{c}^{E}$ as a function of the distance and $h_c$ as $h_{c}^{E}=\frac{(m-1)^2}{(m-1)^2+1}h_c$. According to this relation and in agreement with the results from the finite size system\cite{nishimori05}, it seems that all neighbors are entangled in the vicinity of the quantum critical field. Increasing the magnetic field above the $h_c^{E}$ induces entanglement between the non-nearest neighbor spins. This entanglement that is induced by the external magnetic field, is known as the "magnetic entanglement" and was reported in finite size systems for the first time \cite{nishimori05}. Here we explicitly showed that this phenomenon is also observed in the thermodynamic limit. Furthermore, the concurrence increases to a maximum value before falling to zero at the quantum critical field, $h_c$. In the saturated ferromagnetic phase, $h>h_c$, the non-nearest neighbor spins are not entangled similar to the 1N spins. We also calculated the spin-spin correlation functions $<S_n^{\alpha} S_n^{\alpha}>$ for $\alpha=x, y, z$. We found that the spin-spin correlation functions along the $x$ and $y$ directions dominate  the correlations along the $z$ direction exactly at the critical entangled-field. We have also calculated the concurrence between the 2N spins in finite size chains by using the numerical Lanczos method. The numerical results for chain sizes $N=16, 20,24$ are plotted in the inset of Fig.~\ref{zero temperature}~(a). Although the effect of the level-crossing is seen in the finite size results, there is a region where the 2N spins will be entangled by the magnetic field. This is in agreement with the exact analytical results.

To obtain a better insight into the nature of the entanglement between the non-nearest neighbor spins, we consider the thermal behavior of the entanglement as a general case. Fig.~\ref{hc} shows the temperature dependence of the concurrence between the 2N (Figs.~\ref{hc} (a) and (b)), 3N (Figs.~\ref{hc} (c) and (d)), and 4N (Figs.~\ref{hc} (e) and (f)) spins for different values of the magnetic field $h>h_c^{E}$. Figs.~\ref{hc} (a), (c), and (e) show the results for $h_c^{E}<h\leq(h_c)$. The 2N, 3N, and 4N spins are entangled at zero temperature in this region. By increasing the temperature, the entanglement decreases and vanishes at a critical value $T_c$. The existence of a critical temperature is also reported for concurrence between the 1N spins in this model\cite{shu09}. Here, the critical temperature is a function of the magnetic field strength and increases when the magnetic field approaches the quantum critical value. It is interesting that the critical temperature ($T_c$) has a maximum exactly at $h=h_c=1. 0$. On the other hand, the maximum value of the critical temperature for the 3N spins in this model is almost one-third of the maximum value of the critical temperature for the 2N spins and three times larger than the maximum value of the critical temperature for 4N spins. By comparing Figs.\ref{hc} (a), (c), and (e) we conclude that the entanglement is formed step by step from the 2N, to the 3N and to farther neighbors in the spin chain by reducing the temperature. So the 4N spins are entangled at a lower temperature than the other two pairs.

For the magnetic fields greater than the quantum critical value (Figs.~\ref{hc} (b), (d) and (f)) the 2N, 3N, and 4N spins are not entangled at zero temperature. By increasing the temperature from zero, the spins remain unentangled until the first critical temperature $T_{c_{1}}(h)$ is reached. Once the temperature exceeds $T_{c_{1}}$, the concurrence regains and takes a maximum value, then decreases to zero at the second critical temperature $T_{c_{2}}(h)$. The existence of the second critical temperature is expected, since a sufficiently large thermal fluctuation will destroy all of the classical and quantum correlations. Both of the $T_{c_{1}}$ and $T_{c_{2}}$ values increase by increasing the external magnetic field. We have to mention that the same phenomenon has been observed for the concurrence between the 1N spins in this region of the magnetic field\cite{Mehran14}. The only difference is that the second critical temperature for the concurrence between the 1N spins is almost field-independent. Also note that the entangled region beyond the $h_c$ is very limited for the 3N and 4N spins and diminishes quickly after $h=1.1$ in both cases.

\begin{figure}[t]
\centerline{\psfig{file=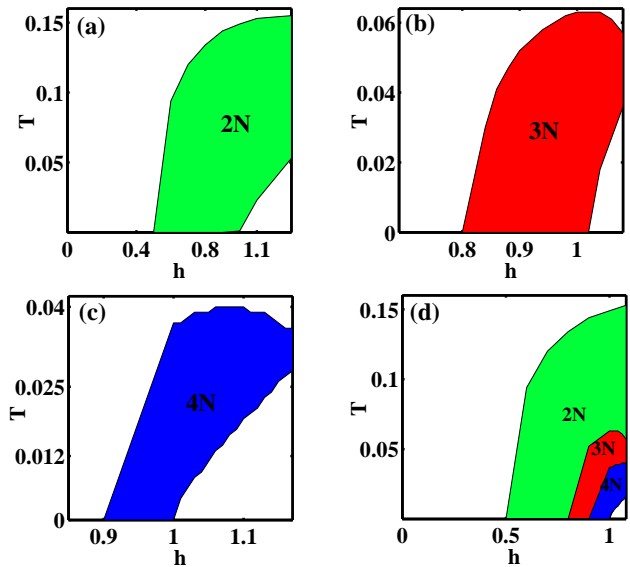,width=3.25in}}
\caption{(color online). Schematic T-h phase diagram for (a) 2N, (b) 3N, (c) 4N spin pairs. Panel (d) shows the overview of entangled region for all three pairs.}
\label{T-h}
\end{figure}

Fig.~\ref{T-h} depicts the T-h phase diagram for (a) 2N, (b) 3N, and (c) 4N pairs. Also Fig.~\ref{T-h} (d) shows an overview of the entangled regions for the 2N, 3N, and 4N pairs. These diagrams signify that there is no entanglement at any temperature interval in the absence of a magnetic field. The borders of each entangled region determine critical temperatures in each magnetic field. It is obvious that $T_{c1}=0$ and $T_{c2}$ has an incremental behavior in $h_c^{E}\leq h \leq h_c$ for all three pairs. For $h>h_c$, however the increasing trend of $T_{c1}$ can be observed for the 2N, 3N, and 4N pairs. $T_{c2}$ shows a different behavior and $T_{c2}$ is almost independent of the magnetic field for the 2N spins, while it decreases with h for the 3N and 4N. According to Fig.~\ref{T-h} (d), one can easily conclude that increasing the magnetic field and decreasing the temperature leads to the induction of entanglement on the farther neighbor spins. Therefore, all pairs are entangled around the $h_c$ at zero temperature.

\begin{figure}[b]
\centerline{\psfig{file=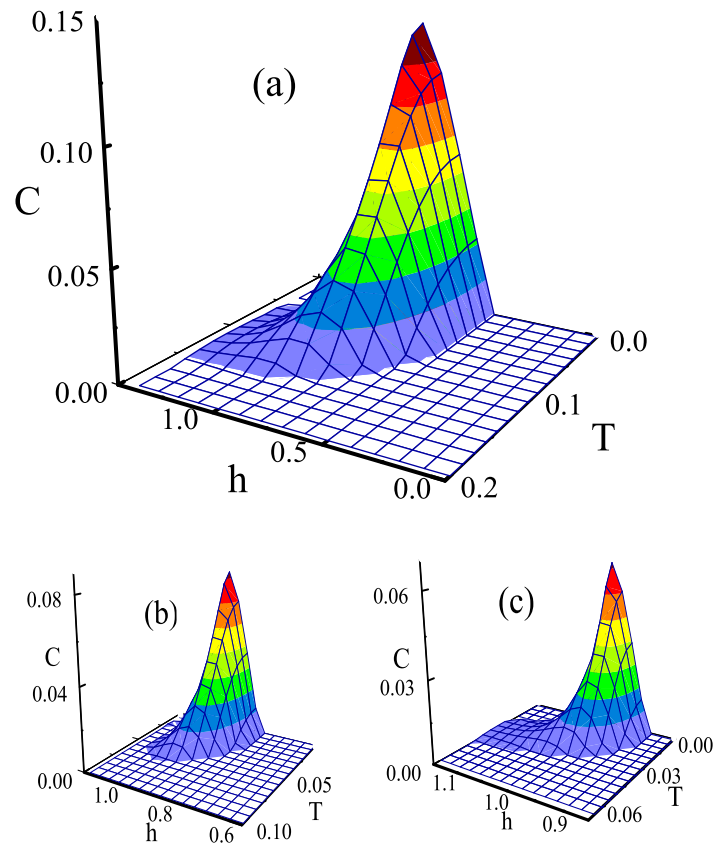,width=3.65in}}
\caption{(color online). The curved surface of the concurrence between (a) 2N, (b) 3N and (c) 4N spins as a function of the temperature and the applied magnetic field for exchange coupling $J=1$.}
\label{C-T-h}
\end{figure}

The complete phase diagram of the concurrence between the 2N, 3N, and 4N spins in the spin-1/2 XY chains in a magnetic field is plotted in Figs.~\ref{C-T-h} (a), (b), and (c) using the exact analytical results.
The concurrence reduces by increasing the temperature in all of the mentioned spin pairs. However, the concurrence magnitude and the thermal interval of the entangled region for the 3N and 4N spins are considerably smaller than the 2N spins.
\section{Conclusion}\label{sec4}

In this work we proposed a generalizable method for calculating the TE between any arbitrary pair of spins in low-dimensional spin chains via the analytical JW transformation. We exactly resolved the concurrence between the 2N, 3N, and 4N spins in the spin-1/2 XY chains in the presence of a magnetic field. We found that all of the spin pairs are entangled at zero temperature where $h_{c}^{E}\leq h \leq h_c$. We also showed that the critical entangled-field, $h_{c}^{E}$, is a fraction of the quantum critical value. The critical entangled-field is $h_c/2$ for 2N spins and $\sim 4h_c/5$ and $9h_c/10$ for the 3N and 4N spins respectively. Furthermore, we studied the temperature dependence of the concurrence between the 2N, 3N, and 4N spins. It is found that the thermal entanglement becomes finite only after $h$ crosses $h_{c}^{E}$ value. When $h$ increases beyond the critical point $h_{c}$, the entanglement is still non-zero and decays gradually which is different from the ground state. The entangled field region is smaller at higher temperatures. Furthermore, unlike the 1N spins entanglement, there is not any unique critical temperature for the non-nearest neighbor spins.

\section{acknowledgments}
The authors wish to thank Islamic Azad University-Langaroud Branch for its financial support based on grant No. 1393-681. We are also thankful to Ahmad K Fumani for reviewing the manuscript.

\vspace{0.3cm}



\begin{thebibliography}{99}


\bibitem{bennet93}
C. H. Bennet, et al., Phys. Rev. Lett. \textbf{70} 1895 (1993).

\bibitem{bouwmeester97}
D. Bouwmeester et al., Nature (London) \textbf{390}, 575 (1997).

\bibitem{gorbachev00}
V. N. Gorbachev and A. I. Trubilko, JETP \textbf{91}, 894 (2000).
\bibitem{bennet92}
C. H. Bennett and S. J. Wiesner, Phys. Rev. Lett. \textbf{69}, 2881 (1992).
\bibitem{hao01}
 J. C. Hao, C. F. Li, and G. C. Guo, Phys. Rev. A \textbf{63}, 054301 (2001).
\bibitem{bennet96}
C. H. Bennet et al., Phys. Rev. A \textbf{54}, 3824 (1996).
\bibitem{hill97}
S. Hill and W. K. Wotters, Phys. Rev. Lett. \textbf{78}, 5022 (1997).
\bibitem{wotters98}
W. K. Wotters, Phys. Rev. Lett. \textbf{80}, 2245, (1998).
\bibitem{vedral97}
V. Vedral et al., Phys. Rev. Lett. \textbf{78}, 2275 (1997).
\bibitem{horodecki98}
M. Horodecki et al., Phys. Rev. Lett. \textbf{80}, 5239 (1998).
\bibitem{rains99}
E. M. Rains, Phys. Rev. A \textbf{60}, 173 (1999); \textbf{60}, 179 (1999).
\bibitem{bayat09}
A. Bayat, S. Bose, Advances in Mathematical Physics \textbf{2010},Article ID 127182, 11 pages (2009).
\bibitem{kastner98}
M. A. Kastner et al., Rev. Mod. Phys. \textbf{70}, 897 (1998).
\bibitem{hase93}
M. Hase, I. Terasaki, and K. Uchinokura, Phys. Rev. Lett. \textbf{70}, 3651 (1993).
\bibitem{gibson04}
B. J. Gibson et al., Physica B \textbf{350},  e253, (2004).
\bibitem{algra78}
H. A. Algra, J. Bartolome, L. J. Dejongh, R. L. Carlin and J. Reedijk, Physical B \textbf{93}, 114 (1978).
\bibitem{blote75}
H. W. J. Blote, Physica (Utrecht) B \textbf{79}, 427 (1975).
\bibitem{kwek09}
L. C. Kwek, Y. Takahashi and K.W. Choo, Journal of Physics: Conference Series \textbf{143}, 012014,(2009).
\bibitem{loss98}
D. Loss and D. P. Divincenzo, Phys. Rev. A \textbf{57}, 120 (1998).
\bibitem{imamoglu99}
Imamoglu et al., Phys. Rev. Lett. \textbf{83}, 4204 (1999).
\bibitem{raussendrof01}
R. Raussendorf and H. J. Briegel, Phys. Rev. Lett. \textbf{86}, 5188 (2001).
\bibitem{wang01}
X. Wang, Phys. Rev. A \textbf{64}, 012313 (2001).
\bibitem{jordan28}
P. Jordan and E. Wigner, Z. Phys. \textbf{47}, 631 (1928).
\bibitem{takahashi99}
M. Takahashi, "Thermodynamics of One-Dimensional Solvable Models", Cambridge University Press, Chapter IV (1999).
\bibitem{Amico02}
L. Amico, R. Fazio, A. Osterloh, V. Vedral, Rev. Mod. Phys. \textbf{80}, 515 (2008).
\bibitem{Nielsen98}
M. A. Nielsen, Ph. D thesis, University of Mexico, quant-ph/0011036 (1998).

\bibitem{Arnesen01}
M. C. Arnesen, S. Bose, and V. Vedral, Phys. Rev. Lett. {\bf 87}, 017901 (2001).

\bibitem{Ghosh03}
S. Ghosh, T. F. Rosenbaum, G. Aeppli, and S. N. Coppersmith,
Nature {\bf 425}, 48 (2003).
\bibitem{Vedral03}
V. Vedral, Nature {\bf 425}, 28 (2003).
\bibitem{Wiesniak05}
M. Wiesniak, V. Vedral, C. Brukner, New J. Phys. {\bf 7}, 258 (2005).
\bibitem{fukuhara15}
T. Fukuhara et al., Phys. Rev. Lett. {\bf 115}, 035302 (2015).
\bibitem{sahling15}
S. Sahling \textit{et al}. Nature Physics, {\bf 11}, 255 (2015).
\bibitem{shu09}
S. Gong and G. Su, Phys. Rev. A. \textbf{80}, 012323, (2009).


\bibitem{arnesen01}
M. C. Arnesen, S. Bose, and V. Vedral, Phys. Rev. Lett. \textbf{87}, 017901 (2001).

\bibitem{Mehran14}
E. Mehran, S. Mahdavifar, R. Jafari, Phys. Rev. A \textbf{89}, 049903 (2014).


\bibitem{nishimori05}
H. Yano and H. Nishimori,Progress of theoretical physics. Supplement \textbf{157}, 164-167, (2005).
\bibitem{osborne02}
T. J. Osborne and M. A. Nielsen, Phys. Rev. A  \textbf{66}, 032110 (2002).
\bibitem{Dillenschneider08}
Raoul Dillenschneider, Phys. Rev. B \textbf{78}, 224413 (2008).

\bibitem{Maziero10}
J. Maziero, H. C. Guzman, L. C. Celeri, M. S. Sarandy, R. M. Serra, Phys. Rev. A \textbf{82}, 012106 (2010).
\bibitem{huang14}
Y. Huang, Phys. Rev. B \textbf{89}, 054410 (2014).
\bibitem{wootters98}
W. K. Wootters, Phys. Rev. Lett. \textbf{80}, 2245-2248 (1998).
\bibitem{amico03}
L. Amico, A. Osterloh, F. Plastina, R. Fazio, . M. Palma, arXiv:quantph/0307048v1, (2003).
\bibitem{osterloh02}
A. Osterloh, L. Amico, G. Falci, and R. Fazio, Nature \textbf{416}, 608 (2002).
\bibitem{nag11}
T. Nag, A. Patra and A. Dutta, journal of statistical mechanics \textbf{08} P08026, (2011).

\bibitem{syljuasen03}
O. F. Syljuasen,  Phys. Rev .A \textbf{68}, 060301 (2003).

\bibitem{lieb61}
E.H. Lieb, T. Schulz and D. Mattis, Ann. Phys. (N.Y.) \textbf{16}, 417 (1961).


\end{thebibliography}
\end{document}